\documentclass[11pt,a4paper]{article}
\pdfoutput=1

\usepackage{jheppub}
\usepackage[rgb]{xcolor}
\usepackage{color}
\usepackage{amsmath,amsfonts,amssymb,mathrsfs,graphicx,bbm,dsfont,booktabs,mathtools,braket,lmodern,tikz,slashed}

\DeclareMathAlphabet{\mathpzc}{OT1}{pzc}{m}{it}

\newcommand{\bea}{\begin{eqnarray}}
\newcommand{\eea}{\end{eqnarray}}
\def\be{\begin{equation}}
\def\ee{\end{equation}}
\newcommand{\bei}{\begin{itemize}}
\newcommand{\eei}{\end{itemize}}
\newcommand{\bee}{\begin{enumerate}}
\newcommand{\eee}{\end{enumerate}}

\newcommand{\alg}[1]{\mathfrak{#1}}

\newcommand{\psu}{\alg{psu}}

\def\ads{{\rm AdS}_5\times {\rm S}^5}

\def\ads{{\rm AdS}_5\times {\rm S}^5}

\def\am{{\rm am}}
\def\am0{{\rm am}_0}

\DeclareMathAlphabet{\mathsc}{T1}{lmr}{m}{scsl}

\expandafter\def\expandafter\bfseries\expandafter{\bfseries\ifmmode\else\boldmath\fi}
\expandafter\def\expandafter\mdseries\expandafter{\mdseries\ifmmode\else\unboldmath\fi}
\expandafter\def\expandafter\normalfont\expandafter{\normalfont\ifmmode\else\unboldmath\fi}

\definecolor{grey}{rgb}{0.4,0.4,0.5}
\definecolor{darkgreen}{rgb}{0,0.5,0}
\definecolor{darkred}{rgb}{0.6,0.0,0}
\definecolor{lightbrown}{rgb}{1,0.9,0.8}
\definecolor{brown}{rgb}{0.6,0.3,0.3}
\definecolor{darkblue}{rgb}{0,0,0.8}
\definecolor{darkmagenta}{rgb}{0.5,0,0.5}

\title{On classical Yang-Baxter based deformations of the $\ads$ superstring}

\author{Stijn J. van Tongeren}

\affiliation{Institut f\"ur Mathematik und Institut f\"ur Physik, Humboldt-Universit\"at zu Berlin, IRIS Geb\"aude, Zum Grossen Windkanal 6, 12489 Berlin, Germany}

\emailAdd{svantongeren@physik.hu-berlin.de}

\abstract{Interesting deformations of $\ads$ such as the gravity dual of noncommutative SYM and Sch\"odinger spacetimes have recently been shown to be integrable. We clarify questions regarding the reality and integrability properties of the associated construction based on R matrices that solve the classical Yang-Baxter equation, and present an overview of manifestly real R matrices associated to the various deformations. We also discuss when these R matrices should correspond to TsT transformations, which not all do, and briefly analyze the symmetries preserved by these deformations, for example finding Schr\"odinger superalgebras that were previously obtained as subalgebras of $\mathfrak{psu}(2,2|4)$. Our results contain a (singular) generalization of an apparently non-TsT deformation of $\ads$, whose status as a string background is an interesting open question.}

\begin{document}

\begin{flushright}\small{HU-EP-15/18\\HU-MATH-15/05}\end{flushright}

\maketitle

\section{Introduction}

Integrability of the string sigma model on $\ads$ has provided important insight into the AdS/CFT correspondence \cite{Maldacena:1997re}. Given the power of the associated techniques \cite{Arutyunov:2009ga,Beisert:2010jr}, considerable effort has been spent on extending them beyond this maximally symmetric example, by deforming the $\ads$ superstring while preserving its integrability \cite{Bena:2003wd}. One important class of such deformations are strings on the Lunin-Maldacena background \cite{Lunin:2005jy} which were generalized and shown to be integrable in \cite{Frolov:2005dj}, see also \cite{Alday:2005ww}.\footnote{See \cite{vanTongeren:2013gva} for a review of the spectral problem in this setting.} More recently, a manifestly integrable deformation of the $\ads$ superstring was constructed in \cite{Delduc:2013qra}, resulting in a quantum deformation of the superconformal symmetry of $\ads$ \cite{Delduc:2014kha}.\footnote{These papers generalize earlier work by Klim\v{c}\'ik \cite{Klimcik:2002zj,Klimcik:2008eq}.} The metric and B field of this model were found in \cite{Arutyunov:2013ega}, a spacetime with very interesting properties \cite{Arutyunov:2013ega,Hoare:2014pna,Arutynov:2014ota,Arutyunov:2014cra,Delduc:2014kha,Kameyama:2014vma,Arutyunov:2014jfa}, but its interpretation in terms of string theory and AdS/CFT remains elusive.\footnote{In the $\mathrm{AdS}_2\times \mathrm{S}^2$ and $\mathrm{AdS}_3\times \mathrm{S}^3$ cases some progress has been made on the supergravity front \cite{Lunin:2014tsa}.} The construction of \cite{Delduc:2013qra} is based on a solution of the so-called modified classical Yang-Baxter equation which essentially has a unique solution \cite{Delduc:2014kha}. However, shortly afterwards it was realized \cite{Kawaguchi:2014qwa} that this construction can be conveniently adapted to yield integrable deformations based on the classical Yang Baxter equation (CYBE), which has a large space of solutions. Moreover, many of these solutions have a nice interpretation in terms of string theory and AdS/CFT, including cases such as the Lunin-Maldacena background \cite{Matsumoto:2014nra}, the gravity dual of noncommutative supersymmetry Yang-Mills theory (SYM) \cite{Matsumoto:2014gwa}, as well as for example certain Schr\"odinger spacetimes \cite{Matsumoto:2015uja}, which were hereby shown to be integrable. However, various confusing statements regarding the reality and integrability properties of these classical Yang-Baxter based deformations, as well as open questions regarding their symmetries and their relation to gravity have arisen in the literature \cite{Kawaguchi:2014qwa,Kawaguchi:2014fca,Matsumoto:2014nra,Matsumoto:2014gwa,Matsumoto:2014ubv,Matsumoto:2015uja,Matsumoto:2014cja}, which we would like to clarify and expand upon in this short paper.

To go into some detail, the $\ads$ superstring action can be represented as a coset sigma model on $\mathrm{PSU}(2,2|4)/(\mathrm{SO}(4,1)\times \mathrm{SO}(5))$ as \cite{Metsaev:1998it}\footnote{Here $h$ is the world sheet metric, $\epsilon^{\tau\sigma}=1$, $A_\alpha = g^{-1} \partial_\alpha g$ with $g\in \mathrm{PSU}(2,2|4)$, $\mathrm{sTr}$ denotes the supertrace, and $d = P_1 + 2 P_2 - P_3$ where the $P_i$ are the projectors onto the $i$th $\mathbb{Z}_4$ graded components of the semi-symmetric space $\mathrm{PSU}(2,2|4)/(\mathrm{SO}(4,1)\times \mathrm{SO}(5))$ (super $\ads$).}
\begin{equation}
\label{eq:AdS5action}
S = -\tfrac{T}{2} \int d\tau d\sigma \tfrac{1}{2}(\sqrt{h} h^{\alpha \beta} -\epsilon^{\alpha \beta}) \mathrm{sTr} (A_\alpha d A_\beta).
\end{equation}
Integrability of this model was established in \cite{Bena:2003wd}. By adapting the arguments of \cite{Delduc:2013qra}, in \cite{Kawaguchi:2014qwa} it was argued that deformations of the $\ads$ superstring action of the form
\begin{equation}
S = -\tfrac{T}{2} \int d\tau d\sigma \tfrac{1}{2}(\sqrt{h} h^{\alpha \beta} -\epsilon^{\alpha \beta}) \mathrm{sTr} (A_\alpha d J_\beta)
\end{equation}
where $J=(1-R_g \circ d)^{-1}(A)$ with $R_g(X)=g^{-1} R(g Xg^{-1}) g$, are also integrable sigma models, provided $R$ is antisymmetric,
\begin{equation}
\mathrm{sTr}(R(m) n) = -\mathrm{sTr}(m R(n)),
\end{equation}
and satisfies the classical Yang-Baxter equation (CYBE)\footnote{The trivial solution $R=0$ gives the undeformed action.}
\begin{equation}
\label{eq:CYBE}
[R(m),R(n)] - R([R(m),n] + [m,R(n)])=0.
\end{equation}
Here $R$ is a linear map from a given Lie (super-)algebra $\mathfrak{g}$ to itself, which can be conveniently represented as
\begin{equation}
R(m)= r_{ij} t_i\, \mathrm{sTr}(t_j m) = \mathrm{sTr}_2(r (1\otimes m))
\end{equation}
for some anti-symmetric matrix $r_{ij}$, where $r=r_{ij} t_i \wedge t_j = \frac{1}{2}r_{ij}  (t_i \otimes t_j -t_j \otimes t_i)$, and the $t_i$ are the generators of $\mathfrak{g}$. In \cite{Kawaguchi:2014qwa} and subsequent works \cite{Kawaguchi:2014fca,Matsumoto:2014gwa,Matsumoto:2014ubv,Matsumoto:2015uja}, this algebra is generically taken to be $\mathfrak{gl}(4|4)$, despite the fact that the parent construction \cite{Delduc:2013qra} suggests $\mathfrak{su}(2,2|4)$. Indeed, it is clear that if we want the deformed action to be real we have to restrict ourselves to $\mathfrak{u}(2,2|4)$.\footnote{Since $A \in \mathfrak{su}(2,2|4)$, the action will be real only when $J$ satisfies the same reality condition, hence $J\in \mathfrak{u}(2,2|4)$, and we see that $R$ must preserve this real form.} Also, upon examination it is clear that the Lax representation of the equations of motion given in \cite{Delduc:2013qra,Kawaguchi:2014qwa}, requires a further restriction to $\mathfrak{su}(2,2|4)$.\footnote{When considering $\mathfrak{gl}(4|4)$ R matrices, we need to consider the derivation of \cite{Delduc:2013qra,Kawaguchi:2014qwa} over $\mathfrak{gl}(4|4)$, not $\mathfrak{sl}(4|4)$, so that we can no longer drop the term in the (matrix valued) equations of motion that is proportional to the identity (see e.g. section 1.2.1 of \cite{Arutyunov:2009ga}). This breaks the equivalence of the equations of motion to the flatness of the proposed Lax pair. We have also checked this explicitly for a $\mathrm{U}(2)$ R matrix deformation of the $\mathrm{S}^2 = \mathrm{SU}(2)/\mathrm{U}(1)$ sigma model. Furthermore, note the related fact that without a restriction to $\mathfrak{su}(2,2|4)$ the action loses its $\mathrm{U}(1)$ gauge symmetry $A \rightarrow A + d \theta 1$, which normally guarantees that the physical degrees of freedom live in $\mathfrak{psu}(2,2|4)$.} Nevertheless, using what appear to be and are referred to as $\mathfrak{sl}(4|4)$ R matrices\footnote{By this we of course mean R matrices whose action is not restricted to $\mathfrak{su}(2,2|4)$.} several interesting deformations of the $\ads$ string action have been constructed that are real, at least at the bosonic level. Furthermore, in \cite{Matsumoto:2014ubv} a deformation based on a $\mathfrak{gl}(4|4)$ R matrix was stated to be integrable, the status of which, given our statements above, requires further investigation. We would like to clarify the situation by showing that \emph{all} real deformations previously considered in this setting, are, or can be, obtained from $\mathfrak{su}(2,2|4)$ R matrices, thereby manifesting the reality of and the existence of Lax pairs for these models. In particular, we give an $\mathfrak{su}(2,2|4)$ rather than $\mathfrak{gl}(4|4)$ R matrix for the new type of deformation considered in \cite{Kawaguchi:2014fca,Matsumoto:2014ubv}.

Most of the deformations considered in this setting, such as the gravity duals of noncommutative SYM and dipole theories, as well as certain Schr\"odinger geometries, are based on commuting symmetry generators, at least heuristically explaining their relation to (null) TsT transformations. The R matrix for the deformation of \cite{Kawaguchi:2014fca,Matsumoto:2014ubv} is not of this type however, which means it should not be possible to obtain it by TsT transformations alone. Given this interesting status, we generalize the associated R matrix, resulting in a singular deformation of $\ads$. We have not verified that this background is a solution of supergravity in general. In one particular case however, this background can be obtained as a TsT transformation in the variables dual to a Sch\"odinger dilation and a perpendicular null translation, indicating that singularities do not distinguish between TsT and non-TsT R matrices, rather that the lack of commutativity of the building blocks of an R matrix does.

We also give a simple description of the symmetries of the deformed theories, in the case of Schr\"odinger spacetimes finding the six and twelve supercharge superalgebras of \cite{Sakaguchi:2008ku},\footnote{This also matches the Killing spinor analysis of \cite{Donos:2009zf}.} up to some subtleties that we explain. For a particular subclass of the deformed backgrounds of \cite{Kawaguchi:2014fca,Matsumoto:2014ubv}, we find a superalgebra with sixteen supercharges. Our generalization of this deformation generically preserves no supersymmetry, but precisely in the case where it corresponds to a TsT transformation, preserves eight supercharges.

In the next section we present an overview of $\mathfrak{su}(2,2|4)$ R matrices corresponding to various deformations studied in the literature.  In section \ref{sec:TsT} we discuss the link between R matrices and TsT transformations, naturally singling out a nonzero class of deformations that does not appear to be of TsT type. In section \ref{sec:symmanalysis} we analyze the symmetries preserved by a given R matrix deformation.

\section{Real CYBE deformations of $\ads$}

Confusion regarding the reality properties of the R matrices used in \cite{Kawaguchi:2014qwa,Kawaguchi:2014fca,Matsumoto:2014nra,Matsumoto:2014gwa,Matsumoto:2014ubv,Matsumoto:2015uja,Matsumoto:2014cja} appears to originate from the (nonstandard but perfectly valid) choice of basis for $\mathfrak{su}(2,2)$ used there. In fact, despite explicit claims to the contrary made in e.g. section 3.4 of \cite{Kawaguchi:2014qwa} and section 2.2 of \cite{Kawaguchi:2014fca}, all R matrices that result in real actions that have been considered in this context \cite{Kawaguchi:2014fca,Kawaguchi:2014qwa,Matsumoto:2014nra,Matsumoto:2014gwa,Matsumoto:2014ubv,Matsumoto:2015uja} preserve the real form $\mathfrak{u}(2,2|4)$. Hence, the actions of these deformed models are manifestly real, including the fermions that have thus far not been investigated. For the readers' convenience, here we will explicitly give an $\mathfrak{su}(2,2|4)$ R matrix for each of the previously considered deformations in the (standard) conventions of \cite{Arutyunov:2009ga}. For completeness we include the already manifestly real R matrix for the Lunin-Maldacena deformation. We discuss the basis of \cite{Kawaguchi:2014qwa,Kawaguchi:2014fca,Matsumoto:2014nra,Matsumoto:2014gwa,Matsumoto:2014ubv,Matsumoto:2015uja,Matsumoto:2014cja}  in appendix \ref{app:nonstandardbasis}.

As reviewed in appendix \ref{app:conventions}, in our conventions the bosonic algebra $\mathfrak{su}(2,2)$ is spanned by the $\gamma^j$ and $[\gamma^k,\gamma^l]$ for indices running from zero to four, while $\mathfrak{su}(4)$ is spanned by $i \gamma^j$ and $[\gamma^k,\gamma^l]$ for indices running from one through five. Note that real exponentials of generators give group elements. We denote the Cartan generators of $\mathfrak{su}(4)$ by $h_j$ where $j$ runs from one to three, which we take to be
\begin{equation}
h_1 = \mathrm{diag}(i,i,-i,-i), \hspace{15pt} h_2 = \mathrm{diag}(i,-i,i,-i), \hspace{15pt} h_3 = \mathrm{diag}(i,-i,-i,i).
\end{equation}
We will focus on the bosonic sector of the models, where we will work with the coset representative
\begin{equation}
g = \left(\begin{array}{cc}g_a & 0 \\ 0 & g_s \end{array}\right),
\end{equation}
with
\begin{equation}
g_a = e^{x_\mu p^\mu} e^{\frac{1}{2} \log z \gamma^4}=(1+x_\mu p^\mu) e^{\frac{1}{2} \log z \gamma^4},
\end{equation}
and
\begin{equation}
g_s = e^{\phi^i h_i} e^{- \frac{\xi}{2} \gamma^1 \gamma^3} e^{\frac{i}{2} \arcsin{r} \gamma^1}.
\end{equation}
Here we have introduced the Poincar\'e translation generators
\begin{equation}
p^\mu = \frac{1}{2}(\gamma^\mu - \gamma^\mu \gamma^4) \in \mathfrak{su}(2,2),
\end{equation}
which are nilpotent
\begin{equation}
p^\mu p^\nu =0.
\end{equation}
With these conventions the undeformed action is the string action of $\ads$ in Poincar\'e coordinates. We will also use the light cone coordinates
\begin{equation}
x^\pm = \frac{x^0 \pm x^3}{\sqrt{2}}.
\end{equation}

\subsection{$\mathfrak{su}(2,2|4)$ R matrices}

\paragraph{The Lunin-Maldacena background} \cite{Lunin:2005jy} for real $\beta$ and its three parameter generalization by Frolov \cite{Frolov:2005dj}, can be found from the $\mathfrak{su}(4)$ R matrix \cite{Matsumoto:2014nra}
\begin{equation}
\label{eq:Rmatrixgammadef}
r = -\epsilon^{ijk} \tfrac{\gamma_i}{8} h_j \wedge h_k,
\end{equation}
the sum over $i$, $j$, and $k$, running from one to three here and elsewhere. The case $\gamma_1=\pm \gamma_2= \pm \gamma_3=\beta$ corresponds to the Lunin-Maldacena background. It is perhaps interesting to note that we cannot enlarge the above R matrix over $\mathfrak{su}(4)$; any solution that contains eqn. \eqref{eq:Rmatrixgammadef} for general $\gamma_i$ is just eqn. \eqref{eq:Rmatrixgammadef}. Enlarging to $\mathfrak{psu}(2,2|4)$ however, amongst others there is an analogous fifteen parameter solution based on its six Cartan generators.

\paragraph{The gravity dual of noncommutative SYM} \cite{Maldacena:1999mh,Hashimoto:1999ut} is obtained from the $\mathfrak{su}(2,2)$ R matrix \cite{Matsumoto:2014gwa}
\begin{equation}
r = a^2 p^0 \wedge p^3 + {a^\prime}^2 p^1 \wedge p^2,
\end{equation}
where $a$ and $a^\prime$ are the parameters used in \cite{Maldacena:1999mh}. Strictly speaking this R matrix gives the Lorentzian continuation of the Euclidean metric presented there in eqn. (2.9), upon also analytically continuing $a^2$, taking us out of $\mathfrak{su}(2,2)$. It is of course possible to generalize this R matrix to a six parameter R matrix based on the independent antisymmetric products of the $p^\mu$.

\paragraph{Dipole type backgrounds} see e.g. \cite{Imeroni:2008cr}, arise from $\mathfrak{su}(2,2)\oplus\mathfrak{su}(4)$ R matrices of the form \cite{Matsumoto:2015uja}
\begin{equation}
\label{eq:Rmatrixdipole}
r = \tfrac{1}{2} p^3 \wedge \alpha^i h_i.
\end{equation}

\paragraph{Schr\"odinger geometries} arise from $\mathfrak{su}(2,2)\oplus\mathfrak{su}(4)$ solutions of the CYBE of the form
\begin{equation}
r = p_- \wedge f
\end{equation}
where $f \in \mathfrak{su}(4)$. One solution that has been studied in particular is \cite{Matsumoto:2015uja}
\begin{equation}
\label{eq:Rmatrixsusyschr}
r = \tfrac{1}{2}p_- \wedge \eta^i h_i.
\end{equation}
This gives a general class of backgrounds found in \cite{Bobev:2009zf}, see also \cite{Bobev:2009mw,Donos:2009zf}.\footnote{The authors appear to present another independent R matrix  in \cite{Matsumoto:2015uja} (eqn. (3.25) there), while it in fact appears to be exactly the same as the above R matrix (eqn. (3.8) there). We hope to hereby avoid further confusion on this part.}

\paragraph{Other generalized scaling geometries} arise from the $\mathfrak{su}(2,2)$ R matrix
\begin{equation}
\label{eq:Rmatrixmysteryname}
r= \tfrac{1}{2} p_- \wedge (a(\gamma^4 + \gamma^0 \gamma^3) + b \gamma^1 \gamma^2).
\end{equation}
The associated backgrounds are equivalent to the ones constructed in \cite{Matsumoto:2014ubv} with $a = c_1 + c_2$ and $b= i(c_1 - c_2)$. For $a=0$ the resulting background corresponds to one originally found in appendix C.1 of \cite{Hubeny:2005qu}. Our R matrix is not quite equivalent to the one used in \cite{Matsumoto:2014ubv} however, as the R matrix of \cite{Matsumoto:2014ubv} contained a $\mathfrak{u}(2,2)$ generator and hence did not strictly speaking manifest integrability, see appendix \ref{app:Rmatrixexample} for details. For future reference, we would like to introduce the following generalization of \eqref{eq:Rmatrixmysteryname}
\begin{equation}
\label{eq:Rmatrixmysterygen}
r= \tfrac{1}{2}  p_- \wedge (a \gamma^4  + b \gamma^1 \gamma^2 + c \gamma^0 \gamma^3),
\end{equation}
which is also a solution of the CYBE. We will come back to these models later.

Finally, a number of other solutions to the CYBE are briefly listed in \cite{Matsumoto:2015uja}.\footnote{Again, note that example 5 in section 3.3 is identical to the subject of section 3.2 there.} We can directly translate them to our basis by following the discussion in appendix \ref{app:nonstandardbasis}.

\subsection*{Solving the CYBE}

Constructively classifying the constant (antisymmetric) solutions of the CYBE over a given Lie (super)algebra is a complicated problem that to our knowledge has not been solved to date.\footnote{See e.g. the introduction of \cite{Gerstenhaber:1997} or \textsection $3.1.\mathrm{D}$ of \cite{Chari} for a discussion of what this would entail in terms of Lie algebra theory.} On the other hand, it is not hard to find various solutions to the CYBE \eqref{eq:CYBE}. In fact, $r=a\wedge b$, where $a$ and $b$ are any two commuting algebra elements, is an obvious general class of solutions.\footnote{We immediately have $[R(m),R(n)] \sim [a,b] =0$, while the other terms are proportional to $\mathrm{sTr}(a[b,x])$ or $\mathrm{sTr}(a[a,x])$ with $x=n$ or $x=m$, which both also clearly vanish.} All previously studied CYBE R matrices are of this type, with the exception of those in eqs. \eqref{eq:Rmatrixmysteryname} and \eqref{eq:Rmatrixmysterygen}, making them particularly interesting. We will come back to this point below.

\section{R matrices and (null) TsT transformations}
\label{sec:TsT}

The Lunin-Maldacena background, the background dual to noncommutative SYM, and also dipole type backgrounds, can all be obtained by TsT transformations, also known as Melvin twists, see e.g. \cite{Imeroni:2008cr}. Moreover, the required TsT transformations precisely involve the pairs of coordinates dual to the pairs of generators appearing in the above R matrices. Similarly, Schr\"odinger geometries are obtained by what is known as a null Melvin twist, which as the name suggests is nothing but the null generalization of a TsT transformation, and the R matrix picture is the same.\footnote{Note that as $g_{+-}\neq0$, a TsT transformation with a shift in $x^-$ (generated by $p_-$) results in a metric deformation of $g_{++}$.} Physically a null TsT transformation (null Melvin twist) consists of (infinite) boosts in addition to pure TsT transformations, but at the formal level a null TsT transformation is nothing but a regular TsT transformation where we shift by a null direction. Let us briefly elaborate on this.

\subsection{Null TsT transformations}

A null Melvin twist can be applied to a background that has a time translation isometry, a spatial translational isometry, and another rotational or translational isometry \cite{Hubeny:2005qu}. Let us call the associated coordinates $t$, $y$ and $z$. The first step of a null Melvin twist is to (Lorentz) boost in the $y$ direction. While in general the background need not have an $\mathrm{SO}(1,1)$ isometry, in our cases it will. Under this boost
\begin{equation}
\left(\begin{array}{c} t \\ y \end{array}\right)\rightarrow \left(\begin{array}{cc} \cosh{\alpha} &  \sinh{\alpha}\\  -\sinh \alpha &  \cosh{\alpha} \end{array} \right) \left(\begin{array}{c} t \\y \end{array}\right)
\end{equation}
After the boost we $T$ dualize in $y$, shift the $z$ field $z \rightarrow z + \beta \tilde{y} $ by the T dual field $\tilde{y}$, and $T$ dualize back. We inversely boost $y$, and finally consider the infinite boost limit $\alpha\rightarrow \infty$, keeping $\beta e^\alpha$ fixed. Clearly in the infinite boost limit the $y$ field involved in the intermediate steps scales as $y_{int} \simeq e^\alpha(-t+y)$ where $t$ and $y$ are the original $t$ and $y$ coordinates. Hence we see that in the limit we are considering, we are doing nothing but a TsT transformation where we shift $z$ by the null direction $x^-$. In fact, as a TsT transformation involves two T dualities, formally a null TsT transformation is really just a TsT transformation where we shift by a null coordinate, and we do not need to strictly follow the above procedure. We refer to a (null) TsT transformation where we T dualize $y$ and shift $z$ as a TsT transformation in $(y,z)$.

Since a TsT transformation is based on two commuting isometries, it is clear that we can associate a solution of the CYBE to any TsT transformation, which then should both produce the same deformation.\footnote{It is possible to explicitly prove that $r=a\wedge b$ with $[a,b]=0$, $\mathrm{sTr}(a^2)\neq0$, and $\mathrm{sTr}(b^2)\neq0$, results in a deformation of the (bosonic) background that is manifestly equivalent to a TsT transformation in $(x_a,x_b)$, where the $x_c$ are the corresponding dual coordinates \cite{hoareunpublished}. Note however that all above deformations with the exception of the Lunin-Maldacena one, involve the nilpotent $p^\mu$.} The converse does not appear to be the case however, as already indicated by the results of \cite{Matsumoto:2014ubv}.

\subsection{Non TsT R matrices?}

It is hard to directly relate an R matrix to a (null) TsT transformation when the associated building blocks do not commute, i.e. $r=c\wedge d$ with $[c,d]\neq0$, since a TsT transformation is inherently associated to two independent (commuting) isometries. An example of this situation is given by the R matrix \eqref{eq:Rmatrixmysteryname} with $a\neq0$, since $[p_-,\gamma^4 + \gamma^0 \gamma^3]\neq0$. Correspondingly it is apparently necessary to use S duality and TsT transformations that involve the sphere in addition to TsT transformations on AdS to reproduce the background \cite{Matsumoto:2014ubv}.

In this light it is interesting to consider the generalized R matrix \eqref{eq:Rmatrixmysterygen}. For $b=0$ the background is given by\footnote{The background can be readily extracted out of the deformed action following e.g. \cite{Arutyunov:2013ega} or \cite{Kawaguchi:2014fca}. We could have presented the associated background for $b\neq0$, but its expression is rather large and uninsightful.}
\begin{equation}
\label{eq:genmysbackground}
\begin{aligned}
ds^2 = \, &  \frac{-2 z^2 dx^+ dx^- + a(a-c)z^{-2}(z dz + \rho d \rho)x^+ dx^+ - a^2(1+\rho^2 z^{-2}) (dx^+)^2}{z^4-(a-c)^2(x^+)^2}\\
& + \frac{d\rho^2 + \rho^2 d\psi^2 + dz^2}{z^2},\\
B = \, &  \frac{(a-c) x^+ dx^- - a( \rho d \rho+ z dz )}{z^4-(a-c)^2(x^+)^2} \wedge dx^+,
\end{aligned}
\end{equation}
where we have used polar coordinates $(\rho,\psi)$ in the $(x_1,x_2)$ plane. This space has a (naked) curvature singularity at $z^2 = |a-c| x^+$, and we should consider the region $z^2 > |a-c| x^+$ to preserve the metric signature. The two terms making up the R matrix \eqref{eq:Rmatrixmysterygen} only commute when $c=-a$. In line with this, we have verified that precisely in this case the above geometry also arises by a TsT transformation in $(y,x^-)$, where $y$ is the coordinate associated with the boost-dilation (Schr\"odinger dilation) $x^+ \rightarrow e^{2\alpha} x^+$, $z\rightarrow e^\alpha z$, $\rho\rightarrow e^\alpha \rho$ generated by $\gamma^4 - \gamma^0 \gamma^3$, thereby explicitly verifying that it is a (singular) solution of supergravity. For other values of $c-a$ there should be no such interpretation, and presumably further duality transformations are required to reproduce the background. It would be interesting to verify that this background is a solution of supergravity in general, especially in light of the conjectures of \cite{Matsumoto:2014nra,Matsumoto:2014cja}. Note that for $a=0$ the deformation looks remarkably elegant
\begin{equation}
\begin{aligned}
ds^2 = \, &  \frac{-2 dx^+ dx^-}{z^2 -c^2 z^{-2} (x^+)^2} + \frac{d\rho^2 + \rho^2 d\psi^2 + dz^2}{z^2} + ds^2_{S^5},\\
B = \, &  \frac{c x^+ z^{-2}}{z^2 -c^2 z^{-2} (x^+)^2} dx^+\wedge dx^-,
\end{aligned}
\end{equation}
but is of course still singular.

\section{(Super)symmetry analysis}
\label{sec:symmanalysis}

The undeformed $\ads$ superstring action has $\mathrm{PSU}(2,2|4)$ symmetry, realized as left multiplication of the coset representative $g$ in the action \eqref{eq:AdS5action} by a constant group element $G$, up to a compensating gauge symmetry $h \in \mathrm{SO}(4,1)\times\mathrm{SO}(5)$
\begin{equation}
G g = g^\prime h.
\end{equation}
This gauge symmetry is preserved under the deformation, and we can ask which part of the global $\mathrm{PSU}(2,2|4)$ symmetry is preserved. Given the construction of the deformed action, this is straightforwardly determined by $R_g$; any transformation $g\rightarrow G g$ that leaves $R_g$ invariant represents an unbroken symmetry of the action. In other words, we want the set of $G$ for which
\begin{equation}
R_G(x)=R(x).
\end{equation}
Translating this statement to the algebra, we want to find the generators $t$ for which
\begin{equation}
\label{eq:Rmatrixsymmetrypreservation}
R([t,x])=[t,R(x)].
\end{equation}
Analyzing this equation for a particular R matrix gives the subalgebra of $\mathrm{psu}(2,2|4)$ that is preserved by a given deformation.

\subsection{Lunin-Maldacena, noncommutative SYM, and dipole theories}

By the above analysis we can readily verify that none of the 32 supercharges of $\mathfrak{psu}(2,2|4)$ are preserved by the R matrix \eqref{eq:Rmatrixgammadef} for the generic $\gamma_i$ deformation, while for $\gamma_1=\pm \gamma_2= \pm \gamma_3=\beta$ we find eight supercharges as appropriate for the $\mathcal{N}=1$ superconformal symmetry of $\beta$ deformed SYM.\footnote{In this paper we always count real supercharges.} Similarly, we can verify that when the rank of $\alpha^i h_i$ is three instead of four, the R matrix \eqref{eq:Rmatrixdipole} preserves four supercharges, while it preserves eight if the rank is two, in agreement with the $\mathcal{N}=1$ and $\mathcal{N}=2$ nonconformal supersymmetry of the dual theories \cite{Bergman:2001rw}. For the dual of noncommutative SYM we find the sixteen supercharges appropriate for a nonconformal $\mathcal{N}=4$ theory \cite{Hashimoto:1999ut}.

\subsection{Sch\"odinger spacetimes}

Given the form of R matrix \eqref{eq:Rmatrixsusyschr} the subalgebras of $\mathfrak{su}(2,2)$ and $\mathfrak{su}(4)$ that satisfy eqn. \eqref{eq:Rmatrixsymmetrypreservation} are given by the centralizer of $p_-$ and $\eta^i h_i$ respectively. Of course, the centralizer of $p_-$ in $\mathfrak{su}(2,2)$ is precisely the Schr\"odinger algebra.\footnote{This is easy to see using the results of \cite{Sakaguchi:2008rx}. The Schr\"odinger algebra can be found there.} For $\mathfrak{su}(4)$ the Cartan subalgebra remains. Now, according to the (more general) Killing spinor analysis of \cite{Bobev:2009mw}, our Schr\"odinger backgrounds of the form $\mathrm{Schr}_5 \times \mathrm{S}^5$ should preserve two Poincar\'e supercharges when $\eta^1 \pm \eta^2 \pm \eta^3 =0$, and four Poincar\'e supercharges when in addition one of the $\eta^i$ vanishes. However, it was subsequently shown \cite{Donos:2009zf} that on special subclasses of solutions which include our backgrounds, the number of supercharges gets enhanced to six, or even twelve. Now our analysis indicates that when the rank of $\eta^i h_i$ is three (i.e. $\eta^1 \pm \eta^2 \pm \eta^3 =0$), six of the 32 supercharges survive, while if the rank is two (i.e. additionally one of the $\eta^i$ vanishes) we get twelve. These six and twelve are split as $2+(2+2)$ and $4+(4+4)$ respectively, where the additional four respectively eight supercharges we find with respect to \cite{Bobev:2009mw} are precisely the extra two respectively four Poincar\'e plus two respectively four conformal supercharges of \cite{Donos:2009zf}. It is also easy to check that these extra conformal supercharges are obtained from the extra Poincar\'e supercharges by the action of the special conformal transformation of the Schr\"odinger algebra \cite{Donos:2009zf}.

In fact, these superalgebras are closely related to those found in \cite{Sakaguchi:2008ku}. According to \cite{Sakaguchi:2008ku}, the Schr\"odinger subalgebras of $\mathcal{N}=2$ and $\mathcal{N}=1$ superconformal subalgebras of $\mathfrak{psu}(2,2|4)$, give superalgebras with twelve supercharges and $\mathfrak{su}(2)^{\oplus2} \oplus \mathfrak{u}(1) \subset \mathfrak{su}(4)$ R symmetry, and six supercharges and $\mathfrak{u}(1)^{\oplus3}$ R symmetry respectively. The superalgebra with six supercharges that we are dealing with here is precisely this latter one, $\mathfrak{u}(1)^{\oplus3}$ being the centralizer of $\eta^i h_i$ in $\mathfrak{su}(4)$ also when $\eta^i h_i$ has rank three. When $ \eta^i h_i$ has rank two on the other hand, our superalgebra with twelve supercharges contains $\mathfrak{su}(2)\oplus \mathfrak{u}(1)^{\oplus2}$ in addition to the Schr\"odinger algebra, instead of the $\mathfrak{su}(2)^{\oplus2}\oplus \mathfrak{u}(1)$ R symmetry of \cite{Sakaguchi:2008ku}. However, looking more closely at the projectors used in \cite{Sakaguchi:2008ku} to derive the $\mathcal{N}=2$ subalgebra, it is clear that one of the $\mathfrak{su}(2)$ algebras actually acts trivially on the supercharges,\footnote{In the notation of \cite{Sakaguchi:2008ku} the generators of the $\mathfrak{so}(4) = \mathfrak{su}(2)^{\oplus2 }$ are realized via (products of) $\gamma$ matrices as $\Gamma^{a^\prime b^\prime}$ with indices running from $5$ to $8$, acting on supercharges from the right. Now their $\mathcal{N}=2$ supercharges satisfy $\tilde{Q} = \tilde{Q}\Gamma^{5678}$ and $\tilde{S} = \tilde{S}\Gamma^{5678}$, meaning that the action of any $\Gamma^{a^\prime b^\prime}$ is equal or opposite to the action of $\Gamma^{c^\prime d^\prime}$, where $c^\prime$ and $d^\prime$ are the elements of $\{5,6,7,8\}$ complementary to $a^\prime$ and $b^\prime$ . Now $\mathfrak{so}(4)$ can be decomposed as $\mathfrak{su}(2) \oplus \mathfrak{su}(2)$ precisely via such linear combinations of complementary pairs of $\Gamma^{a^\prime b^\prime}$, and we can easily check that one of the two $\mathfrak{su}(2)$s then acts trivially on the $\mathcal{N}=2$ supercharges.} and we would reinterpret their twelve supercharge Schr\"odinger superalgebra as the direct sum of a superalgebra with $R$ symmetry $\mathfrak{su}(2) \oplus \mathfrak{u}(1) = \mathfrak{u}(2)$ and a separate $\mathfrak{su}(2)$ factor. Here we simply have this minimal Schr\"odinger superalgebra extended by a central $\mathfrak{u}(1)$ instead of $\mathfrak{su}(2)$. Similar statements of course apply to the $\mathfrak{u}(1)^{\oplus3}$ of the six supercharge superalgebra, where the minimal superalgebra in fact has only $\mathfrak{u}(1)$ R symmetry,\footnote{In \cite{Sakaguchi:2008ku} all fermions have the same charge under each of the three $\mathfrak{u}(1)$s, hence we can decouple two (linear combinations) of the $\mathfrak{u}(1)$s.} and it is not hard to check that the six supercharges we find from eqn. \eqref{eq:Rmatrixsymmetrypreservation} are indeed charged under only one independent linear combination of our Cartan generators. We believe this answers the question raised in \cite{Matsumoto:2015uja} regarding the relation between R matrices and Schr\"odinger superalgebras.

\subsection{Generalized scaling spacetimes}

Repeating our analysis for the background \eqref{eq:genmysbackground}, we find that it preserves eight supercharges for $a=-c$, sixteen for $a=c$, and none otherwise.\footnote{Strictly speaking, we are finding supersymmetries of the string sigma model that do not always have to arise from supersymmetries of the supergravity background, see e.g. \cite{Duff:1997qz,Duff:1998us,Arutyunov:2014cra,Arutyunov:2014jfa}.} These eight resp. sixteen supercharges form one resp. two fundamental representations of $\mathfrak{su}(4)$, which itself is untouched. Taking $b\neq0$ breaks all supersymmetry regardless of the values of $a$ and $c$. Regarding the remaining bosonic symmetries, note that these geometries do not have full Schr\"odinger symmetry  for nonzero $a$ or $c$, generically only preserving null translations, rotations in the $(x_1,x_2)$ plane and the Schr\"odinger dilation mentioned earlier, as can also be easily seen from the background.

\section{Conclusions}

We hope to have clarified the reality and manifest integrability properties of various R matrices used to construct integrable deformations of the $\ads$ superstring, aspects of the (lack of a) link between R matrices and TsT transformations, as well as the symmetries of our R matrix deformed models. We believe that one of the most interesting remaining open questions is the classification of solutions to the classical Yang-Baxter equation, which may not be entirely out of reach for the specific case of $\mathfrak{psu}(2,2|4)$ and its lower dimensional analogues. For example, the CYBE has a two parameter family of solutions over $\mathfrak{su}(1,1)$ and no solution over $\mathfrak{su}(2)$. This two parameter $\mathfrak{su}(1,1)$ R matrix is of the form $r \sim a \sigma_1 \wedge \sigma_2 + a \cos \theta (i \sigma_3) \wedge \sigma_1 + a \sin \theta (i \sigma_3) \wedge \sigma_2$. Of course we are only really interested in R matrices up to global transformations. In this case we can change $\theta$ by an $\mathfrak{su}(1,1)$ transformation (a time translation in the coordinates below), and we might as well set it to zero. Unfortunately this R matrix does not appear to yield a very pleasant deformation of $\mathrm{AdS}_2$, giving a metric dependent on global time $t$\footnote{We work with the coordinates of e.g. \cite{Arutyunov:2013ega}.}
\begin{equation*}
ds^2 = \left(1-a^2\left(2\rho^2 + (1+2\cos(2t)) (1+\rho^2) - 4 \rho \sqrt{1+\rho^2} \cos t\right)\right)^{-1}ds^2_{AdS_2},
\end{equation*}
preserving none of the $\mathrm{SU}(1,1)$ symmetry of $\mathrm{AdS}_2$. In Poincar\'e coordinates this metric depends on time as well. Of course, many of the nice deformations we considered in the previous sections were simpler, based on commuting elements, inherently requiring higher rank algebras. A related interesting direction would be to investigate and if possible prove the equivalence of R matrices built on commuting elements to (null) TsT transformations. Even more relevant is the question whether any solution of the CYBE corresponds to a string background as conjectured in \cite{Matsumoto:2014nra,Matsumoto:2014cja}, and if so, what the link is between an R matrix built out of noncommuting elements, and e.g. S duality in addition to TsT transformations.\footnote{S duality does not generically preserve integrability.} One way to gain further insight here would be to investigate the supergravity embedding of the generalized background \eqref{eq:genmysbackground}, and others of its kind. It might also be interesting to investigate R matrices involving the fermionic generators of $\mathfrak{psu}(2,2|4)$; thus far only one such solution has been given but not further investigated \cite{Kawaguchi:2014qwa}.\footnote{The parts of an R matrix that involve fermionic generators by default do not deform the bosonic background, and presumably correspond to turning on fermions in the background. The physical interpretation in particular of R matrices that mix bosonic and fermionic generators would be quite peculiar.}

Finally, many of the powerful tools associated with the integrability of the $\ads$ superstring \cite{Beisert:2010jr}, arise in the so-called exact S matrix approach \cite{Arutyunov:2009ga}. This approach relies on the `BMN' light cone gauge that requires isometries in global AdS time and one of the angles on the sphere. Correspondingly, while tools such as the thermodynamic Bethe ansatz can be extended from $\ads$ \cite{Arutyunov:2009zu,Arutyunov:2009ur,Bombardelli:2009ns,Gromov:2009bc} to quantum deformations \cite{Arutynov:2014ota,Arutyunov:2012zt,Arutyunov:2012ai},\footnote{See \cite{Beisert:2008tw,Hoare:2011wr,Arutyunov:2013ega} for the quantum deformed exact S matrix this is built upon.} it is not immediately clear whether this can be done for most of the deformations described here. All deformations here, with the exception of the Lunin-Maldacena-Frolov deformation,\footnote{The thermodynamic Bethe ansatz can indeed be extended to this case \cite{Arutyunov:2010gu,Ahn:2011xq,deLeeuw:2012hp}, see also \cite{vanTongeren:2013gva}. Interestingly, note also that there is a clear link between the R matrix \eqref{eq:Rmatrixgammadef} and the twist of the twisted exact S matrix picture of \cite{Ahn:2010ws}, which presumably extends beyond this case provided the form of the twist is compatible with the exact S matrix.} are naturally suited to AdS in Poincar\'e coordinates, and result in a metric with time dependence in (naive) global coordinates. It would be great to understand whether we can extend the exact S matrix approach to e.g. the dual of noncommutative SYM or Schr\"odinger spacetimes.


\section*{Acknowledgements}

I would like to thank N. Bobev, B. Hoare, T. Matsumoto, and K. Yoshida for discussions, and G. Arutyunov, B. Hoare, T. Matsumoto, and K. Yoshida for comments on the paper. ST is supported by LT. This work was supported by the Einstein Foundation Berlin in the framework of the research project "Gravitation and High Energy Physics", and acknowledges further support from the People Programme (Marie Curie Actions) of the European Union's Seventh Framework Programme FP7/2007-2013/ under REA Grant Agreement No 317089 (GATIS).

\appendix

\section{Appendices}

\subsection{Algebra conventions}
\label{app:conventions}

In this paper we are mainly concerned with the bosonic subalgebra $\mathfrak{su}(2,2)\oplus\mathfrak{su}(4)$ of $\mathfrak{\psu}(2,2|4)$. For details on the material presented here, as well as its supersymmetric extension, we refer to the pedagogical review \cite{Arutyunov:2009ga}. We will only briefly list the facts we need, beginning with the $\gamma$ matrices
\begin{equation}
\begin{aligned}
&\gamma^0 = i \sigma_3 \otimes \sigma_0,  &\gamma^1 = \sigma_2 \otimes \sigma_2, &&\gamma^2 = -\sigma_2 \otimes \sigma_1, \\ &\gamma^3 = \sigma_1 \otimes \sigma_0, &\gamma^4 = \sigma_2 \otimes \sigma_3, &&\gamma^5 = -i \gamma^0,
\end{aligned}
\end{equation}
where $\sigma_0 = 1_{2\times2}$ and the remaining $\sigma_i$ are the Pauli matrices. With these matrices the generators of $\mathfrak{so}(4,1)$ in the spinor representation are given by $n^{ij} = \frac{1}{4} [\gamma^i,\gamma^j]$ where the indices run from zero to four, while for $\mathfrak{so}(5)$ we can give the same construction with indices running from one to five. The algebra $\mathfrak{su}(2,2)$ is spanned by these generators of $\mathfrak{so}(4,1)$ together with the $\gamma^i$ for $i=0,\ldots,4$, while $\mathfrak{su}(4)$ is spanned by the combination of $\mathfrak{so}(5)$ and $i\gamma^j$ for $j=1,\ldots,5$. Concretely, these generators satisfy
\begin{equation}
m^\dagger \gamma^5 + \gamma^5 m = 0
\end{equation}
for $m\in \mathfrak{su}(2,2)$, and
\begin{equation}
n^\dagger+ n = 0
\end{equation}
for $n \in \mathfrak{su}(4)$. This means that we are dealing with the canonical group metric $\gamma^5 = \mathrm{diag}(1,1,-1,-1)$ for $\mathrm{SU}(2,2)$, and $e^{\alpha n}$ and $e^{\alpha m}$ give group elements for real $\alpha$.

The generator $\Omega$ of the $\mathbb{Z}_4$ automorphism of $\mathfrak{\psu}(2,2|4)$ acts on these bosonic subalgebras as
\begin{equation}
\Omega(m) = -K m^t K,
\end{equation}
where $K=-\gamma^2\gamma^4$, which leaves the subalgebras $\mathfrak{so}(4,1)$ and $\mathfrak{so}(5)$ invariant.

\subsection{A different basis}
\label{app:nonstandardbasis}

The conventions of \cite{Kawaguchi:2014qwa,Kawaguchi:2014fca,Matsumoto:2014nra,Matsumoto:2014gwa,Matsumoto:2014ubv,Matsumoto:2015uja,Matsumoto:2014cja} are based on $\gamma$ matrices introduced there as\footnote{As these conventions differ from the above, the statement in appendix A of \cite{Kawaguchi:2014fca} that they basically follow \cite{Arutyunov:2009ga} is somewhat confusing.}
\begin{equation}
\begin{aligned}
&\tilde{\gamma}^0 = i \sigma_2 \otimes \sigma_3,  &\tilde{\gamma}^1 = \sigma_2 \otimes \sigma_2, &&\tilde{\gamma}^2 = -\sigma_2 \otimes \sigma_1, \\ &\tilde{\gamma}^3 = \sigma_1 \otimes \sigma_0, &\tilde{\gamma}^5 = i \tilde{\gamma}_1 \tilde{\gamma}_2 \tilde{\gamma}_3 \tilde{\gamma}_0, &&\tilde{\gamma}^4 = -i \tilde{\gamma}^0.
\end{aligned}
\end{equation}
$\mathfrak{su}(2,2)$ is then defined by imposing the reality condition
\begin{equation}
\label{eq:altrealitycond}
\tilde{n}^\dagger \tilde{\gamma}^0 + \tilde{\gamma}^0 \tilde{n} = 0,
\end{equation}
on elements of $\mathfrak{sl}(4)$ \cite{Matsumoto:2015uja}. With this definition, $\mathfrak{su}(2,2)$ is spanned by $\{\tilde{\gamma}^j, \frac{1}{4} [\tilde{\gamma}^k,\tilde{\gamma}^l]\}$, where indices run over $0,1,2,3,$ and $5$. Since the eigenvalues of $\tilde{\gamma}^0$ are $\pm i$, strictly speaking we should define not $\tilde{\gamma}^0$ but e.g. $\tilde{\gamma}^4$ as the group metric, but otherwise this is a perfectly valid choice of basis. Real exponentials of algebra elements again give group elements. Note however that the `Cartan generators' introduced in e.g. appendix A of  \cite{Matsumoto:2015uja}, are not all elements of $\mathfrak{su}(2,2)$; $\tilde{h}_1$ there does not respect the real form, being defined as $\tilde{h}_1 \equiv i \tilde{\gamma}^1 \tilde{\gamma}^2$.

Also here we can use the $\tilde{\gamma}$ matrices to provide a basis for $\mathfrak{su}(4)$, now given by the $i \tilde{\gamma}^j$ and $\frac{1}{4} [\tilde{\gamma}^k,\tilde{\gamma}^l]$ for indices running from one through five.\footnote{There appears to be a typographical error in \cite{Matsumoto:2015uja} where it is stated that $\mathfrak{su}(4)$ is given by the real span of the $\tilde{\gamma}^j$ and $\frac{1}{4} [\tilde{\gamma}^k,\tilde{\gamma}^l]\}$; $\mathfrak{su}(4)$ reality requires an $i$ on the $\tilde{\gamma}$s.}

Of course it is possible to relate the basis described in this section to the one of \cite{Arutyunov:2009ga} by a $\mathrm{GL}(16)$ transformation. In fact, $\tilde{\gamma}^i = \gamma^i$ for $i=1,2,3$, and all we need to do is identify $\tilde{\gamma}^4 = \gamma^5$, in line with the above discussion of the group metric. Note that $\tilde{\gamma}^4 \rightarrow \gamma^5$ cannot be achieved by a $\mathrm{GL}(4)$ transformation on the matrices themselves, if we wish to preserve the remaining $\gamma^i$.

\subsection{An example R matrix}
\label{app:Rmatrixexample}

Our prototypical example of a Jordanian R matrix\footnote{Jordanian R matrices are antisymmetric solutions of the CYBE that are nilpotent; $R^k(m)=0$ for $k\geq 3$.} written in the above nonstandard basis will be
\begin{equation}
\label{eq:Rjordexample}
r = e_{24} \wedge (c_1 e_{22} - c_2 e_{44})
\end{equation}
as considered in \cite{Matsumoto:2014ubv} and for $c_1=c_2 \in \mathbb{R}$ in \cite{Kawaguchi:2014fca}. Here the $e_{ij}$ denote the matrix unities, with a one in row $i$, column $j$. In \cite{Kawaguchi:2014qwa} it is explicitly stated that Jordanian R matrices do not preserve the real form $\mathfrak{su}(2,2|4)$. This was repeated in \cite{Kawaguchi:2014fca} for this particular R matrix, while noting that in the case $c_1 = c_2 \in \mathbb{R}$ it nevertheless results in a real action. However, provided $c_2^* = c_1$, the above R matrix clearly is built out of generators that satisfy the reality condition \eqref{eq:altrealitycond}, and therefore it does preserve the real form $\mathfrak{u}(2,2|4)$. This explains the observation of  \cite{Matsumoto:2014ubv} that $c_1 = c_2^*$ is a sufficient condition for reality of the action.

Still, $c_1 e_{22} - c_2 e_{44}$ can have nonzero trace, meaning \eqref{eq:Rjordexample} is generically a $\mathfrak{u}(2,2)$ R matrix, obscuring its status with regard to integrability. Still, we claim to reproduce the same background via the $\mathfrak{su}(2,2)$ R matrix \eqref{eq:Rmatrixmysteryname}. To explain this, we need to discuss the inversion of the operator $1-R_g \circ d$ in the action. By construction, the domain of the inverse operator is $\mathfrak{(p)su}(2,2|4)$, and by the supertrace only the $\mathfrak{su}(2,2|4)$ projection of its range appears in the action. Hence, the only place where one can distinguish between a $\mathfrak{u}(2,2|4)$ and an $\mathfrak{su}(2,2|4)$ R matrix, is in the space over which we invert $1-R_g \circ d$. As the projection of a $\mathfrak{u}(2,2|4)$ R matrix onto $\mathfrak{su}(2,2|4)$ is also a solution of the CYBE over $\mathfrak{su}(2,2|4)$, we can (attempt to) invert such an R matrix over both $\mathfrak{u}(2,2|4)$ and $\mathfrak{su}(2,2|4)$. The outcome will generically be different however; consider for example the $(1,1)$ entry of
\begin{equation}
\left(\begin{array}{cc} 1 & r \\ -r & 1 \end{array}\right)^{-1}
\end{equation}
and compare it to $1^{-1}$. Now, if we invert the R matrix over $\mathfrak{su}(2,2|4)$, no $\mathfrak{u}(2,2|4)$ aspect of the R matrix ever enters the problem, and we should simply consider the projection of the R matrix onto $\mathfrak{su}(2,2|4)$ as our R matrix. Of course, were we to invert over $\mathfrak{u}(2,2|4)$, we would generically find problems in the matching of the Lax pair construction with the equations of motion. It is easy to check this explicitly for the $\mathrm{S}^2$ sigma model.\footnote{The CYBE has no solutions over $\mathfrak{su}(2)$, but it has one over $\mathfrak{u}(2)$. As it currently stands then, the q-deformation of \cite{Delduc:2013fga} based on the modified CYBE is the unique real and manifestly integrable deformation of the $S^2$ sigma model in this (R matrix) setting.} Going over the derivation in \cite{Kawaguchi:2014fca,Matsumoto:2014ubv} we see that they in fact invert over $\mathfrak{su}(2,2|4)$, so that they are really using an $\mathfrak{su}(2,2|4)$ R matrix disguised as a $\mathfrak{u}(2,2|4)$ one. Our R matrix \eqref{eq:Rmatrixmysteryname} is the $\mathfrak{su}(2,2|4)$ projection of this R matrix translated to our basis.

\bibliographystyle{JHEP}

\bibliography{Stijnsbibfile}

\end{document}